\documentstyle[12pt]{article}
\begin{document}
\baselineskip=0.7cm
\newcommand{\EQ}{\begin{equation}}
\newcommand{\EN}{\end{equation}}
\newcommand{\EQA}{\begin{eqnarray}}
\newcommand{\EQN}{\end{eqnarray}}
\newcommand{\e}{{\rm e}}
\newcommand{\Sp}{{\rm Sp}}
\renewcommand{\theequation}{\arabic{section}.\arabic{equation}}
\newcommand{\Tr}{{\rm Tr}}
\renewcommand{\thesection}{\arabic{section}.}
\renewcommand{\thesubsection}{\arabic{section}.\arabic{subsection}}
\makeatletter
\def\section{\@startsection{section}{1}{\z@}{-3.5ex plus -1ex minus
 -.2ex}{2.3ex plus .2ex}{\large}}
\def\subsection{\@startsection{subsection}{2}{\z@}{-3.25ex plus -1ex minus
 -.2ex}{1.5ex plus .2ex}{\normalsize\it}}
\def\appendix{
\par
\setcounter{section}{0}
\setcounter{subsection}{0}
\def\thesection{\Alph{section}}}
\makeatother
\def\thefootnote{\fnsymbol{footnote}}
\begin{flushright}
hep-th/9912255\\
UT-KOMABA/99-20\\
December 1999
\end{flushright}
\vspace{1cm}
\begin{center}
\Large
Spontaneously Broken Space-Time Supersymmetry \\
in Open String Theory without GSO Projection 
 
\vspace{1cm}
\normalsize
{\sc Tamiaki Yoneya}
\footnote{
e-mail address:\ \ {\tt tam@hep1.c.u-tokyo.ac.jp}}
\\
\vspace{0.3cm}

{\it Institute of Physics, University of Tokyo\\
Komaba, Meguro-ku, 153 Tokyo}

\vspace{1cm}
Abstract
\end{center}

We argue that orientable open string theory 
without GSO projection has N=2 space-time supersymmetry 
in a spontaneously broken phase.  The arguments are 
presented  both in Neveu-Schwarz-Ramond and Green-Schwarz 
formulations. 
The formal but explicit supersymmetry 
transformation law for string field is given in 
the framework of Witten's  open string field theory. 
Our results  support a 
fundamental assumption which lies behind the 
topological construction of stable 
D-branes starting from the 
unstable systems of D9-branes.

\newpage
\section{Introduction}
One of the crucial observations in the development of 
superstring theory has been the discovery of 
the so-called GSO projection 
\cite{gso} for realizing the 
space-time supersymmetry.  The GSO projection plays dual roles in superstring theory: It eliminates 
tachyons and simultaneously 
makes the number of space-time bosonic 
and fermionic physical excitations equal at each mass level. 
 In the most recent developments of string theory 
associated with various string dualities and 
D-branes, the space-time supersymmetry and GSO projection 
are playing indispensable roles. For example, the 
BPS property of D-branes is a direct consequence of 
the GSO projection. 

On the other hand, from the early days of 
string theory, the condensation of tachyon 
in the theories without GSO projection 
has been among several fundamental 
dynamical questions which have long 
been resisting our exploration.  
Recently, in connection with the problem of constructing stable D-branes in unstable D-brane systems, 
interests on the tachyon condensation \cite{sen}
have been revived.  The general 
topological characterization of D-branes has been given based on 
the K-theoretical interpretation \cite{witten}\cite{horava}  of unstable D-brane systems. 
The systems of unstable D-branes 
are realized as various open string theories 
with mixed (namely, 
combined ordinary and opposite) GSO projections  or without 
the projection at all.  
Undoubtedly,  formulating the dynamics  of such systems 
must be among future cornerstones  
for the development of non-perturbative string/M theory 
towards the ultimate unified theory.  

A simple and fundamental example among
 systems without supersymmetry is 
 unstable D9-branes \cite{horava} in type IIA  
theory. This system, in a low energy approximation,  
may be approximated by 
10 dimensional Yang-Mills theory coupled with 
tachyon and with vector-like Majorana fermions. 
If the stable D-branes are to be realized 
as topological excitations with 
BPS property resulting from the condensation of tachyon, 
the system should have a hidden supersymmetry, perhaps 
as a nonlinear realization, in a spontaneously broken phase.  
This is a basic  assumption behind recent descriptions  
of unstable D-brane systems using tachyons.  
For example, approaches to world-volume actions \cite{sen1} for unstable D-branes and the condensation of tachyon 
\cite{sen2} have been proposed 
along this line. 
However,   no concrete 
evidence justifying this assumption from the viewpoint of 
the dynamics of open strings has been given.  
The purpose of the present work is to fill this gap 
by providing some 
positive  arguments in support of this conjecture. 
Since any possible models for the dynamics 
of such systems already contain 
the string length 
parameter $\ell_s =\sqrt{\alpha'} $ as the inverse tachyon mass,  any local-field theory approximation cannot be 
regarded as a systematic way of dealing with 
the problem except for discussing certain 
long-distance behaviors. 
In general, we have to take into account all 
mass levels from the outset.   

After presenting some elementary arguments 
for spontaneously broken
 symmetries of string theory in next  two sections, we 
 investigate in section 4 the case of 
space-time supersymmetry first in the 
NSR formalism.  Then we argue that the 
same results are obtained in 
the Green-Schwarz formalism as well. 
In section 5,  we proceed to 
investigate the formal supersymmetry transformation law 
 in the framework of Witten's open superstring field theory.   
We conclude in section 6 by 
making some remarks.  
We will also briefly discuss how the linear 
supersymmery could be recovered in certain 
nonperturbative vacua. 

\section{Evidence for spontaneously broken space-time supersymmetry}
\setcounter{equation}{0}
Let us start from  
collecting 
some pieces of qualitative evidence for the existence of 
space-time supersymmetry without 
the GSO projection in order to motivate our 
discussions in later sections.  
Since, without the projection, the number of 
physical particle degrees of freedom 
are not equal between fermions and bosons at 
fixed mass level, we naturally expect that the 
tachyonic vacuum must break 
supersymmetry spontaneously even if 
supersymmetry is actually a symmetry of the theory.  
Is there any evidence for this?  
First of all, there must exist Goldstone fermions. 
Since the Ramond sector indeed contains massless fermions 
as ground state, this is possible.  Let us first 
briefly study the coupling of tachyon with 
massless fermions 
to get some flavor 
on how such an interpretation should look like. 

Without the GSO projection, the massless fermions  
appear as a pair of opposite chiralities. 
Let them be $\psi_{\pm} \, \, (\gamma_{11}\psi_{\pm}=\pm
\psi_{\pm})$, and tachyon be $T$. Then the 
field equation for fermions 
are of the form
\EQ
\gamma\cdot \partial \psi_{\pm} 
= g T\psi_{\mp} + O(T^2 \psi_{\pm}) + \cdots  
\EN
where we have suppressed the coupling with the 
massless gauge fields. 
The natural candidate for the supercurrent is 
\EQ
s^{\mu}_{\pm} \equiv  \gamma^{\mu}\psi_{\pm} 
+ ag(\partial^{\mu}T)\psi_{\mp} 
+ bgT\partial^{\mu}\psi_{\mp} 
+ O(g^2T^2 \psi_{\pm}) +\cdots .
\EN 
Note that we have taken into account the conservation of G-parity  in Neveu-Schwarz-Ramond 
open string theory in writing down these expressions,  
and $a, b$ are constants 
to be determined.  $g$ is the coupling constant of open strings. 
The usual string coupling of closed strings $g_s$ is 
proportional to the square of $g$. 
The first term which is linear in the massless fermions  
is the signature of Goldstone fermions and of the nonlinear 
realization of supersymmetry.  
Taking the divergence of the 
would-be supercurrent leads to 
\EQ
\partial \cdot s_{\pm} 
= \gamma\cdot \partial \, \psi_{\pm}
+ag\Bigl(
\partial^2 T \, \psi_{\mp} + (\partial_{\mu}T) \partial^{\mu}\psi_{\mp}
\Bigr)
+bg\Bigl(
(\partial_{\mu}T)\partial^{\mu}\psi_{\mp} 
+T \partial^2 \psi_{\mp}
\Bigr)
+\cdots .
\EN
Now using the tachyon equation of motion 
\EQ
\partial^2 T = -{1\over \alpha'}T + O(g^2T^3) + \cdots 
\EN
together with the fermion equation of motion, we find that 
the choice $a=\alpha'=-b$ leads to the  
conservation of the supercurrent up to the order $O(T^2)$
\EQ
\partial\cdot s_{\pm} = O(g^2T^2) + \cdots
\EN
By requiring that the fermion equation is consistent 
with the conservation of supercurrent 
to the next order $O(g^2T^2)$, we find that 
the order $O(g^2T^2\psi)$ term in the fermion 
equation and the current are 
\EQ
\alpha' (1+2c) g^2(T\partial_{\mu}T)\gamma^{\mu}\psi_{\pm}, 
\quad 
-\alpha'c g^2 T^2 \gamma^{\mu} \psi_{\pm}
\label{ordertwocorrection}
\EN
respectively, where $c$ is an undetermined constant. 
Of course, we have to include the gauge fields and 
all the higher excitations to achieve a 
really consistent construction as well as more complicated nonlinear 
interaction terms. 
However, this argument serves 
as a guide on what we should expect. For example, 
the result (\ref{ordertwocorrection}) suggests 
that the condensation of tachyon such that 
$g^2c T^2=1/\alpha'$ recovers the linear 
supersymmetry by eliminating the inhomogeneous 
 term (and hence also the original kinetic terms of them 
in the action) in the supercurrent. 

The existence of spontaneously broken supersymmetry 
when all massive levels are 
taken into account 
is suggested by the following observation, 
which, to the author's knowledge, has never been 
mentioned in the literature although it is implicitly 
contained in a familiar formula. 
For a wide class of local field theory models  with 
supersymmetry, it is well known that 
the tree-level mass levels after the spontaneous 
breakdown of supersymmetry satisfy 
the trace formula \cite{massformula} for the mass square operator $M^2$ 
$((-1)^F =1 $ or$ -1$ for bosons and fermions, respectively)
\EQ
\Tr\Bigl(
(-1)^F M^2 
\Bigr)=0
\label{tracemsquare}
\EN
where the trace $\Tr(\cdot)$ is taken over all physical  one-particle states. 
This of course reduces to ${\rm tr}((-1)^F)=0$, the trace "tr" now 
being taken at 
each {\it fixed} mass level, if the 
supersymmetry is realized linearly without spontaneous breaking. 

Let us check whether similar relations 
among different mass levels are valid  in open 
string theory with tachyon.  
In string theory containing states of infinitely large 
masses, it is natural to define regularized traces 
corresponding to this quantity by  
\EQ
A_n \equiv \lim_{\tau\rightarrow 0}
\Tr\Bigl(
(-1)^F M^{2n}\e^{-\tau M^2}
\Bigr) =(-1)^n\lim_{\tau\rightarrow 0}{d^n\over d\tau^n}
Z(\tau) ,
\EN 
\EQ
Z(\tau) = \Tr\Bigl(
(-1)^F\e^{-\tau M^2}
\Bigr)  .
\EN
In open NSR string theory, we have 
\EQ
Z(\tau)=\Tr_{NS}\Bigl(
q^{2N_{NS}-1}
\Bigr)
-\Tr_{R}\Bigl(
q^{2N_{R}}
\Bigr)
\label{zett}
\EN
where $q=\e^{-\tau/2\alpha'}$ and 
$N_{NS}, N_{R}$ are the level operators  
for the NS sector and R sector, respectively. 
\EQA
N_{NS}&=&\sum_{n=1}^{\infty}\alpha_{-n}\alpha_n
+\sum_{r=1/2}^{\infty}rb_{-r}b_{r}, \\
N_R&=&\sum_{n=1}^{\infty}\alpha_{-n}\alpha_n
+\sum_{n=1}^{\infty}nd_{-n}d_n .
\EQN
Note that there is no GSO projection in taking trace in 
(\ref{zett}) and we have suppressed the transverse 
spatial indices for the creation-annihilation operators. 
Since the GSO projected pieces in this trace 
cancel between the NS and R sector owing to the 
well-known Jacobi's `{\it abstrusa}' identity, 
it is sufficient to retain only the contribution from the 
oppositely projected pieces.  We then have 
\EQ
Z(\tau)={1\over \prod_{n=1}^{\infty}(1-q^{2n})^8}
\Bigl({1\over 2q}
\prod_{n=1}^{\infty}(1+q^{2n-1})^8 +
{1\over 2q}\prod_{n=1}^{\infty}(1-q^{2n-1})^8
-{1\over 2}16 
\prod_{n=1}^{\infty}(1+q^{2n})^8
\Bigr) .
\EN
The $+$ sign in front of the second term in this expression 
indicates the opposite GSO projection. Because of this,  
a negative power term $q^{-1}$ is not canceled,
 corresponding to the existence of 
tachyon in the NS sector.  To study the short time 
behavior $\tau\rightarrow 0$, we perform Jacobi transformation $q=\e^{-\pi \sigma} 
\rightarrow q=\e^{-\pi/\sigma} \, \, (\sigma=\tau/2\pi\alpha') $ using 
theta function identities, and obtain 
\EQA
Z(\tau)=&&\hspace{-0.4cm}
{\sigma^4\over 2}{1\over 
\prod_{n=1}^{\infty}(1-e^{-2n\pi/\sigma})^8}\times \\ 
\nonumber 
&&\hspace{-2cm}\Bigl(
 \e^{{\pi\over \sigma}}\prod_{n=1}^{\infty}
(1+e^{-(2n-1)\pi/\sigma})^8+16 \prod_{n=1}^{\infty}
(1+\e^{-2n\pi/\sigma})^8 
- \e^{{\pi\over \sigma}}\prod_{n=1}(1-\e^{-(2n-1)\pi/\sigma})^8
\Bigr) .
\EQN
The asymptotic behavior  in the 
short time limit $\sigma \rightarrow 0$ is 
\EQ
Z(\tau) \rightarrow 
{\sigma^4\over 2}\Bigl(16 + 16 + (256+ 256) \e^{-2\pi/\sigma} 
+ O(\e^{-4\pi/\sigma})\Bigr) .
\EN
If the ordinary GSO projection were used, 
we would instead have 
$16-16$, $256-256$, and so on, 
  in this expression. As is well known 
\cite{pol},  
the ``$16+16$"-contribution corresponds to the 
massless NS-NS + R-R fields in the closed-string channel. 
If we sum over the GSO projected and oppositely projected 
parts, only the states in the NS-NS sector 
remain, showing that there is 
no R-R charges exchanged in the $t$-channel. 
 
Now the above formula clearly shows that 
\EQ
A_{n}=0  , \quad (n\ne 4). 
\label{traceformula}
\EN 
Thus the graded (regularized) traces  over all mass levels 
are almost all zero except for $n=4$ in 
NSR open string theory. This should 
be regarded as a generalization of the 
mass formula (\ref{tracemsquare}) for spontaneously broken supersymmetry 
in local field theory to open superstring theory with the 
tachyonic vacuum.   The fact that 
$A_n$'s vanish for arbitrary large $n$ suggests that 
the possible nonlinear 
supersymmetry could only be realized 
when all mass levels are taken into account. 

We note that the result (\ref{traceformula})  is essentially 
a direct consequence of the $s$-channel-$t$-channel duality 
and  that, in the closed-string ($t$) channel, 
the ordinary GSO projection is automatic if we include both NS and R sectors in the 
open string ($s$) channel.  Thus there is no 
problem of nonlocality on the world sheet 
even if we do not make the 
GSO projection in the open-string channel. 
Otherwise, we could not have 
assumed the unprojected (or oppositely projected) 
open string theories consistently as  describing the unstable D-branes in superstring theory. 

\section{Nonlinear symmetries in string theory}
\setcounter{equation}{0}
Before proceeding to further discussions 
on spontaneously broken supersymmetry, let us here 
recall how the nonlinear symmetry transformations  
and associated spontaneous symmetry breaking can be 
realized in string theory in its general world-sheet formulation.  
In the present paper, 
for brevity of  terminology, we call any symmetry 
transformation with inhomogeneous term 
`nonlinear' symmetry. 
The ordinary gauge transformation and 
also the general coordinate transformation 
are typical examples for this phenomena. 

One of the characteristic features of space-time symmetries 
exhibited in  physical scattering amplitudes of string theory 
is that they are 
always associated with the corresponding currents 
on the world sheet. 
Take as an instructive example the gauge transformation 
of an open string state for which the massless 
vector state just plays the role of gauge field. 
In the world-sheet picture, this corresponds to the 
existence of the following world-sheet current
\EQ
j_{\lambda}^a(\xi) \equiv \epsilon^{ab} \partial_b \lambda (x(\xi)) ,
\EN
where we have only considered the bosonic part 
for simplicity and $\lambda(x)$ is an arbitrary 
massless external scalar field. $(\xi^0, \xi^1)=(\tau, \sigma)$ 
are the world-sheet coordinates ($0\leq \sigma \leq \pi$). 
Here and in what follows the vector indices $a, b, \ldots$ 
correspond to world sheet and those 
$\mu, \nu, \dots $  to target space-time. 
All world-sheet local operators should be understood as 
normal-order products.  To make the following 
manipulations well-defined after the normal ordering 
we have to impose the usual on-shell conditions. 
Now it is obvious that although the 
current is conserved locally $\partial_a j_A^a=0$ 
on the world sheet, the charge 
$q_{\lambda}=\int d\sigma j^0_{\lambda}$ is not conserved 
for Neumann open strings. 
\EQ
{d q_{\lambda}(\tau) \over d\tau}=\pi p^{\mu}(\tau,\sigma) 
\partial_{\mu}\lambda(x(\tau, \sigma))\Big|^{\sigma =\pi}_{\sigma=0}
\label{u1charge}
\EN
where $p^{\mu}={1\over \pi}\partial_0 x$ is the 
canonical momentum on the world sheet 
(our unit is $\alpha'=1/2$).  Note that the violation of 
charge conservation is 
proportional to the vertex operator 
for the massless vector state.\footnote{
Note that in the covariant operator language, the violation 
of charge conservation is measured by the 
commutator with the Virasoro  (or BRST) 
operators. To the present author's knowledge, 
the derivation of nonlinear gauge and space-time symmetries 
in string theory 
from this point of view was first discussed in ref. \cite{yoneold}. 
} If some of the 
directions obey the Dirichlet condition, 
the violation of charge conservation only comes from 
Neumann directions. 
On the other hand, the charge itself is 
given as 
\EQ
q_{\lambda}(\tau) 
=\lambda(x(\xi))\Big|^{\sigma=\pi}_{\sigma=0}. 
\EN
Namely, the charge generates the phase 
gauge transformation of the string wave function 
\EQ
\Psi (\tau) \rightarrow \e^{-iq_{\lambda}(\tau) }
\Psi (\tau)=\e^{i(\lambda(x(\tau, 0))-\lambda(x(\tau, \pi)))}
\Psi (\tau) . 
\EN
The equation (\ref{u1charge}) shows 
that we have to make a 
shift of the on-shell external gauge 
field $(\partial^2 A_{\mu}=0, 
\partial_{\mu}A^{\mu}=0)$ 
corresponding to the massless vector excitations 
\EQ
\delta A_{\mu}(x) = \partial_{\mu}\lambda (x) , 
\EN
in order to compensate the violation of the conservation law 
at the string boundaries.  

This trivial example tells us that in string theory 
in general we cannot exclude the violation of 
symmetry occurring at the string boundary, {\it provided that
 the violation is proportional to 
physical vertex operators} and hence is compensated by the 
insertion of the vertex operators.  
Actually, to include the symmetry associated 
with the closed-string states, 
this must be further 
generalized to the situation \cite{yoneold2} 
where the current is 
not conserved even locally on the world sheet, but 
the violation is proportional to 
vertex operators of closed strings.  
The general coordinate symmetry and gauge 
transformation corresponding to the antisymmetric 
NS-NS tensor belong to this latter class. 
The currents corresponding to them are 
\EQ
T_{a}= v_{\mu}(x)\partial_a x^{\mu}
\EN
and 
\EQ
B^{a}=\epsilon^{ab}u_{\mu}(x)\partial_b x^{\mu} ,
\EN
respectively, where $u_{\mu}, v_{\nu}$ are 
arbitrary on-shell divergenceless-and-massless vector fields. 
Their divergences, by the world-sheet equation of 
motion, are 
\EQ
\partial_a T^{a}= {1\over 2}
(\partial_{\nu} v_{\mu}+\partial_{\mu}v_{\nu})
\partial^a x^{\nu}\partial_a x^{\mu} ,
\EN
\EQ
\partial_a B^a = {1\over 2} 
(\partial_{\nu} v_{\mu}-\partial_{\mu}v_{\nu})
\epsilon^{ab}\partial_a x^{\mu}\partial_bx^{\nu} .
\EN
Since the divergences are proportional to the 
on-shell gauge transformation 
\[
\delta h_{\mu\nu}=\partial_{\nu} v_{\mu}+\partial_{\mu}v_{\nu}
, \quad 
\delta b_{\mu\nu} =\partial_{\nu} v_{\mu}-\partial_{\mu}v_{\nu}
\]
of 
the graviton and NS-NS antisymmetric tensor 
and hence are compensated by the 
on-shell gauge transformation of these fields, 
the above currents give the symmetry of string theory. 
Indeed the charges are nothing but the generator 
of general coordinate (in the sense of 
target space-time) and string-shape dependent phase 
transformations ($|\psi\rangle \rightarrow \exp i\int d\sigma b_{\mu}(x(\sigma))
\partial_{\sigma}x^{\mu}(\sigma)|\psi\rangle$), respectively. We also remark that 
in the case of open strings 
the current corresponding to general coordinate 
transformation has no extra violation at Neumann boundaries, 
while the conservation 
of the NS-NS antisymmetric tensor-gauge charge 
is violated at Neumann boundaries. 
It is a common knowledge that
 the latter violation is  
compensated by the shift of the vector gauge field 
$\delta A_{\mu} =\pi u_{\mu}$ which couples at the boundary.   This is the origin of a now familiar 
relationship between Dirac-Born-Infeld action 
and the non-commutative Yang-Mills theory \cite{seiwitten}. 

The reason why we come to this seemingly self-evident 
 digression is 
that these examples contain the case of the 
symmetries which are spontaneously broken 
in the ordinary perturbative vacua. 
Take the first example for definiteness. If we consider 
the gauge function 
$\lambda(x) = c_{\mu}x^{\mu}$ which is linear in the coordinate, the gauge transformation of the gauge 
field is nothing but the constant shift of the 
field
\EQ
\delta A_{\mu}=c_{\mu} ,
\EN
which is spontaneously broken in 
perturbative vacua.  It is well known that 
the massless gauge bosons,  including 
photon and graviton, on the perturbative vacuum 
can be regarded as the Goldstone bosons  
associated with this phenomena.  
One important remark here is that 
since the constant vector $c_{\mu}$ is not 
constrained, all components of the constant vector 
is in a sense physical. Thus at vanishing momentum 
the number of physical states are increased. This is  
nothing but the phenomenon of `discrete' physical states 
and is the signature of the existence of spontaneously 
broken symmetries associated with local space-time symmetries.  
Remember that the gauge transformations with gauge 
functions that do not vanish at infinity can in general be 
spontaneously broken. 
Almost the same apply to the cases of 
constant metric and constant antisymmetric field $B_{\mu\nu}$.  
In the former, the transformation is 
a general GL(10, R) coordinate transformation, 
while in the latter the transformation corresponds 
to a linear vector gauge function $b_{\mu}=B_{\mu\nu}x^{\nu}$ 
and a linear shift of vector gauge field $A_{\mu}=B_{\mu\nu}x^{\nu}$.   

\section{Nonlinear supersymmetry in NSR and GS open string theories}
\setcounter{equation}{0}
We now study the space-time supersymmetry in light of 
the above observations.  Our goal is to show that 
the massless fermion states of open string theory 
can be regarded as the Goldstone fermions 
associated with the spontaneously broken supersymmetry 
in the tachyonic vacua. 

\subsection{The NSR formalism}

In the NSR formalism, the 
charge corresponding to space-time supersymmetry 
is the integral of fermion emission vertex at 
zero-momentum. Using the standard result 
of \cite{fms}, it is 
\EQ
Q_{\alpha}=\oint{dz\over 2\pi i}V_{\alpha}^F ,
\label{scharge}
\EN
\EQ
V_{\alpha}^F(z)=\lim_{k\rightarrow 0}S_{\alpha}(z)
\e^{-\phi(z)/2}\e^{ikx(z)} ,
\EN
where we are using the $-1/2$ picture and 
$\alpha$ of the world-sheet spin operator $S_{\alpha}$ is the 32 components SO(10) space-time spinor 
index. The scalar field $\phi(z)$ corresponds to the  
standard bosonized representation of the superghosts 
$(\gamma, \beta) = (\e^{\phi}\eta, \e^{-\phi}\partial\xi) $. 
In the open-string notation, the components of 
the world-sheet current $s^a(\tau, \sigma)$ corresponding to 
the supercharge (\ref{scharge}) ($Q_{\alpha}=
\int_0^{\pi}d\sigma s_{\alpha}^0$) is 
\EQ
s_{\alpha}^0(\tau,\sigma) = {1\over 2}(S_{\alpha}\e^{-\phi/2}(\tau-\sigma) 
+ S_{\alpha}\e^{-\phi/2}(\tau+\sigma)) ,
\EN
 \EQ
s_{\alpha}^1(\tau,\sigma) = {1\over 2}(S_{\alpha}\e^{-\phi/2}(\tau-\sigma) 
- S_{\alpha}\e^{-\phi/2}(\tau+\sigma)) .
\EN
Here we have formally shifted  to the convention of the world-sheet Minkowski metric for comparison with the 
convention of the previous section. However, in considering 
the OPE on the world-sheet, we always assume the 
Euclidean convention. 

The usual argument \cite{fms} for the GSO projection comes from the 
multi-valuedness of the spin operator. Namely, the operator 
product expansion with the world-sheet Neveu-Schwarz field 
$\psi^{\mu}(z)$ is 
\EQ
\psi^{\mu}(z)S_{\alpha}(w) \sim (z-w)^{-1/2}\gamma_{\alpha\beta}S^{\beta}(w) + \cdots , \, \, 
 etc .
\label{psisproduct}
\EN
In closed-string theory, this forces us to project out 
the states with odd G-parity in the NS sector, and 
simultaneously the half of states classified by the 
generalized chirality operator in the R-sector as 
required by the operator product (\ref{psisproduct}), 
in order to ensure the integrability of the fermion-emission vertex operators 
on the Riemann surface. 
As is well known, after including the ghost contributions,  
the G-parity and the generalized chirality operators 
which we call 'fermionic G-parity', are given,  
in terms of the oscillators, by 
\EQ
G=-(-1)^{\sum_{r=1/2}^{\infty}(d_{-r}d_r + \beta_{-r}\gamma_r -\gamma_{-r}\beta_r)} ,
\EN 
\EQ
\Gamma=\gamma_{11}(-1)^{\sum_{n=1}^{\infty}
d_{-n}d_n + \beta_{-n}\gamma_n - \gamma_{-n}
\beta_n + \beta_0\gamma_0} .
\EN

In open string theory, however, the projection is not 
completely mandatory since the open string vertex operators are inserted only at the boundary of world sheets. 
The on-shell closed string states which appear by the 
factorization of the world-sheet Riemann surface 
with such vertex operators for open string states 
inserted at the boundaries 
are automatically GSO projected, provided 
we include both NS and R states, as 
exemplified by the partition function 
treated in section 2.  There appear no tachyonic 
states in closed-string channel. 
 
Now if we include the oppositely GSO-projected states 
in open-string theory, the world-sheet supercharge can still 
formally be defined by 
\EQ
Q_{\alpha}(\tau)=\int_{0}^{\pi}d\sigma \, s_{\alpha}^0(\tau, \sigma) , 
\label{susychargeintegral}
\EN
by choosing a particular Riemann sheet. 
Then, the charge  is not in general conserved, 
but the current is conserved locally on the world sheet and possible violation of charge conservation comes only from the 
open-string boundaries.  
\EQ
{d Q_{\alpha}(\tau)\over d\tau}
=-\int _0^{\pi}d\sigma \, \partial_1 s_{\alpha}^1(\tau, \sigma) =-S_{\alpha}\e^{-\phi/2}(\tau-\pi) 
+ S_{\alpha}\e^{-\phi/2}(\tau+\pi) .
\EN
The right hand side does not vanish in general since 
the operator 
$S_{\alpha}\e^{-\phi/2}$ can be double-valued. In that case, 
$S_{\alpha}\e^{-\phi/2}(\tau-\pi)- S_{\alpha}\e^{-\phi/2}(\tau+\pi) \rightarrow 2S_{\alpha}\e^{-\phi/2}(\tau-\pi)$ which is itself proportional 
to fermion emission vertex at the string end point ($\sigma=\pi$)  with 
zero momentum.  The cases where double-valuedness occurs   
can be classified in two classes: First,  
when the supercurrent acts on bosonic states (NS sector),  
the double-valuedness occurs if and only if the states are oppositely GSO projected, 
independently on the fermionic G-parity $\Gamma$ of the 
supercurrent. Secondly, when the supercurrent 
acts on fermionic states (R-sector), it occurs if and only if 
the fermionic G-parities of the supercurrent and 
that of the states are mutually opposite. In the 
first case, the fermionic G-parity of the states obtained after the action of the supercurrent is opposite to that of the supercurrent, whereas in the second case the bosonic states 
are always oppositely projected. This means that 
these two cases are consistent. Namely, both corresponds to 
3-point vertices with 2 fermions and 1 boson 
in which the fermionic G-parities of two fermion lines are  
flipped with the emission of one odd G-parity boson.  

The violation of supercharge conservation, being 
compensated by the condensation of constant 
massless fermion states,  can be 
interpreted in  the same way 
as in the case of the spontaneous breaking of the 
gauge symmetry corresponding to the constant shift 
of the vector gauge fields. Thus, the supersymmetry 
is broken spontaneously.  
In terms of oscillator modes,  the supersymmetry 
transformation mixes all mass levels and cannot in general 
be truncated into  multiplets of 
finite dimensions. The reason is that the 
natural moding of the supercurrent 
density for double valued states are 
half-integer expansion.   
But the supercharge is still defined by the 
integral (\ref{susychargeintegral}) with no 
half-integer exponential factor. 
If we try to represent the symmetry transformation 
in terms of massless excitations by eliminating 
all the massive levels, the symmetry transformation is 
literally nonlinearly realized. 

We also note that 
at zero-momentum 
there is no Dirac equation for the wave functions 
which are  contracted with the spinor index of the 
supercurrent. Hence, 
the number of degrees for the constant shift of the 
massless fermion fields is doubled, comparing with the 
on-shell physical states obtained as the zero-momentum 
limit of the massless fermions with non-zero momentum: $8+8 
\rightarrow 16+ 16$.   The corresponding 
Goldstone fermions are just the vector-like 
massless fermion with both chiralities. 
In conclusion, there exists a nonlinearly 
realized $N=2$  (IIA-type) space-time supersymmetry in 10 space-time dimensions. 

\subsection{The GS formalism}

We next study how the above situation is 
described in the Green-Schwarz formalism,  
since it is usually more suitable for realizing the space-time 
supersymmetry.  
In order to do this, it is useful to first reconsider 
the partition function 
which we have discussed in section 2. 
Jacobi's  identity leads to the following identity 
for the partition function of the oppositely 
projected states. 
\EQA
&&{1\over \prod_{n=1}^{\infty}(1-q^{2n})^8}
\Bigl({1\over 2q}
\prod_{n=1}^{\infty}(1+q^{2n-1})^8 +
{1\over 2q}\prod_{n=1}^{\infty}(1-q^{2n-1})^8
-{1\over 2}16 
\prod_{n=1}^{\infty}(1+q^{2n})^8
\Bigr) \nonumber \\
&&\hspace{4cm}= {1\over q}{\prod_{n=1}^{\infty}(1-q^{2n-1})^8 \over 
\prod_{n=1}^{\infty}(1-q^{2n})^8} .
\label{partitionanti}
\EQN
What is the interpretation of the right hand side?
The answer is 
that the expression 
 is nothing but the partition function 
for the GS open string theory with the `{\it wrong}' 
boundary condition, which violates the manifest ({\it i.e.} linear) 
space-time supersymmetry but is a perfectly valid condition 
from the viewpoint of the pure world-sheet field theory. 
Indeed the prefactor $q^{-1}$ 
corresponds to the Casimir energy that is not 
cancelled between bosons and fermions because of the 
half-integer moding of the spinor coordinates forced 
by the `wrong' (namely, anti-periodic) boundary condition.  
The infinite product factor  in the numerator is nothing but the 
contribution from the excitation modes of 
the spinor coordinates, with the minus sign being 
the space-time fermion grading.  
Note that there is no zero modes for the 
spinor coordinates, due to the anti-periodic boundary condition. 
The conservation 
of the world-sheet current corresponding to 
the space-time supersymmetry is only violated at 
the open-string boundaries, so that we can again 
interpret this as an indication of the spontaneously 
broken supersymmetry. 

Let us discuss more details. 
Unfortunately, in the GS formalism, the  
quantization is only practicable in the light-cone gauge . Because of this,  
the following discussion is somewhat limited, but   
it provides reasonable evidence that the same results as 
in the NSR formalism are valid in the GS formalism. 
For some aspects, the latter is more transparent 
in treating the NS and R sectors on equal footing.  

Let us briefly recapitulate the ordinary 
formulation of the light-cone GS string 
\cite{gs}\cite{gsw}. We will mostly follow the 
convention of \cite{gsw}. 
The light-cone gauge 
action  is 
\EQ
A_{light-cone}=\int d\tau d\sigma\Bigr(
-{1\over 2\pi}\partial_aX^i \partial^aX^i 
+ {i\over 4\pi}\overline{S}\gamma^- \rho^a\partial_a S
\Bigr)
\EN
Here space-time vector index  $i$ runs only over the 
transverse directions, and $S$ denotes two  Majorana-Weyl 
space-time spinors as fermion  
fields on the world sheet satisfying the 
light-cone gauge condition,   
\EQ
\gamma^+ S=0 . 
\label{lightconeconds}
\EN
The two-component index $n$  of $S_{\alpha n}$ with respect to world-sheet,  
being  acted upon by the two-dimensional 
Dirac matrices $(\rho^0, \rho^1) $, is 
behaving as a two-dimensional spinor because of the 
gauge fixing of the local 
$\kappa$ symmetry of the covariant action 
\cite{gs2} 
and hence $\overline{S}=S\gamma^0\rho^0$. 
The (space-time) supersymmetry transformation 
is 
\EQA
\delta X^i &=&(p^+)^{-1/2}\overline{\epsilon}\gamma^i S ,\\
\delta S &=& i(p^+)^{-1/2}\gamma_-\gamma_{\mu}\rho\cdot \partial X^{\mu}
\epsilon .
\EQN
In the transformation of the spinor $S$,  
$X^{-}$ does not appear and 
$\partial_0 X^+=p^+$, due to the light-cone gauge conditions. 
The world-sheet supercurrents corresponding to this 
transformation is 
\EQ
s_{\alpha}^a=
\gamma_{\mu}(\rho\cdot \partial X^{\mu})\rho^aS_{\alpha} .
\EN
Note that the currents do not satisfy the light-cone 
condition (\ref{lightconeconds}).   

We have to impose the boundary condition 
that identifies the two (world-sheet) 
components of the field $S$ 
in order to ensure the world-sheet energy-momentum 
conservation at the boundary. 
The usual condition which keeps the space-time 
supersymmetry manifestly is 
\EQ
S_{\alpha 1}(\tau, 0)=S_{\alpha 2}(\tau, 0) , \quad 
S_{\alpha 1}(\tau, \pi)=S_{\alpha 2}(\tau, \pi) .
\label{pericond}
\EN
Thus the number of independent components of spinor coordinates 
are 8 which belongs to the irreducible spinor 
representation ${\bf 8}_s $ of SO(8) and 
are represented by a single periodic holomorphic spinor 
coordinate field $S_A(\tau-\sigma) =
S_A(\tau-\sigma+2\pi) $ defined as 
\EQ
S_A(\tau-\sigma) = S_{\alpha 1}(\tau, \sigma)\, \,  {\rm for} 
 \, \, 0\le \sigma \le\pi, \quad 
S_A(\tau-\sigma) = S_{\alpha 2}(\tau, 2\pi-\sigma)\, \, 
{\rm for}\, \,  \pi \le \sigma \le 2\pi .
\label{holospinor}
\EN
We denote the SO(8) spinor indices corresponding to 
representations $({\bf 8}_s, {\bf 8}_c) $ by 
$(A,\dot{A})$.  Note that an SO(10) chiral 
spinor with 16
components now consists of a direct sum of two 
irreducible spinor representations ${\bf 8}_s +{\bf 8}_c $, 
namely, $\alpha \sim (A, \dot{A})$, of which 
${\bf 8}_c \,  (\sim \dot{A})$ are dependent 
components.  Similarly, an opposite chirality 
(in the sense of 10 dimensions) spinor consists 
of the representations ${\bf 8}_c +{\bf 8}_s$, $\alpha \sim (\dot{A}, A)$ 
with ${\bf 8}_s\sim A$ being now the on-shell dependent components 
 with non-zero momentum.    

The independent world-sheet supercharges 
corresponding to the above supersymmetry 
 transformation are then 
\EQA
Q_A(\tau)&=&(2p^+)^{1/2}\int_0^{\pi} d\sigma
(S_A(\tau-\sigma)+S_A(\tau+\sigma)) ,\\
Q_{\dot{A}}(\tau) &=&(p^+)^{-1/2}\gamma^i_{B\dot{A}}
\int_0^{\pi} d\sigma(p^i(\tau-\sigma)S_B(\tau-\sigma) 
+p^i(\tau+\sigma)S_B(\tau+\sigma))
\EQN
where $p^i(\tau-\sigma)$ is the usual transverse orbital 
momentum operator with mode expansion 
$p^i(\tau-\sigma)=\sum_n\alpha_n \e^{-in(\tau-\sigma)} $,  
again momentarily returning to world-sheet Minkowski 
notation.  $\gamma^i_{A\dot{B}}$ are the $8\times 8$ 
SO(8) matrices satisfying $\gamma^i_{A\dot{B}} \gamma^{j}_{\dot{B}C} 
+\gamma^j_{A\dot{B}} \gamma^{i}_{\dot{B}C}=2\delta^{ij}\delta_{AC}$ 
with $\gamma^i_{\dot{B}A}$ being the transpose 
of $\gamma^i_{A\dot{B}}$. 
The world-sheet supercurrents are correspondingly decomposed into 
\EQA
s^0_A(\tau)&=&(2p^+)^{1/2} 
(S_A(\tau-\sigma)+S_A(\tau+\sigma)) ,\\
s^1_A(\tau)&=&(2p^+)^{1/2} 
(S_A(\tau-\sigma)-S_A(\tau+\sigma)) ,\\
s^0_{\dot{A}}(\tau) &=&(p^+)^{-1/2}\gamma^i_{B\dot{A}}
 (p^i(\tau-\sigma)S_B(\tau-\sigma) 
+p^i(\tau+\sigma)S_B(\tau+\sigma)) , \\
s^1_{\dot{A}}(\tau) &=&(p^+)^{-1/2}\gamma^i_{B\dot{A}}
 (p^i(\tau-\sigma)S_B(\tau-\sigma) 
-p^i(\tau+\sigma)S_B(\tau+\sigma)) .
\EQN
The supersymmetry charges are conserved since the 
spinor coordinates $S_A(\tau-\sigma)$ are periodic. 
The spectrum of this theory precisely reproduces 
that of the GSO projected states of NSR open strings. 
There exists a 10 dimensional 
 space-time symmetry ($N=1$) with 16 supercharges. 

We now consider what happens if the boundary condition is replaced by another possibility, instead of (\ref{pericond}),  
\EQ
S_{\alpha 1}(\tau, 0)=-S_{\alpha 2}(\tau, 0) , \quad 
S_{\alpha 1}(\tau, \pi)=S_{\alpha 2}(\tau, \pi) .
\label{antipericond}
\EN
In this case, the spinor 
coordinates defined again by (\ref{holospinor}) 
are anti-periodic $S_A(\tau, \sigma+2\pi)=-
S_A(\tau, \sigma)$ .  
The mass square operator is 
\EQ
\alpha' M^2_{GS-}= 
\sum_{n=1}^{\infty}\alpha^i_{-n}\alpha^i_n 
+ \sum_{r=1/2}^{\infty} r S_{-rA}S_{rA} -{1\over 2} ,
\EN
which leads to the partition function (with fermion grading) 
(\ref{partitionanti}) and hence exactly 
matches with the spectrum of the oppositely 
GSO projected states, as already discussed in the 
beginning of this subsection.  
 Here we have used the 
standard notation: The modes of the 
antiperiodic spinor coordinates labeled by 
half integers are $S_{rA}$.  
The uncanceled zero-point energy $-1/2$ can 
be derived either from  Lorentz invariance or the 
standard rule from the zeta-function regularization.  
Thus, the ground state is the 
scalar tachyon. The first excited states are 
$u_{A-}(k) S_{-1/2A}|0\rangle $ which are massless fermion states $k^2=0$.  The subscript $-$ indicates that the 
fermion state should be interpreted as the 
$-$ chirality state with respect to SO(9,1), in 
contrast to the massless fermion states in the usual GS string 
which have $+$ chirality.   
The space-time handedness of the massless 
fermion states  can be interpreted to be opposite to those 
appeared in the case of periodic boundary condition.  
This is due to the fact that the massless fermion states 
now come from the non-zero modes of the spinor coordinates. 
If the massless vector states of the ordinary GS open string is 
denoted by $|i\rangle $, the massless fermion states 
of the ordinary GSO projection, 
which come from the spinor zero modes 
of the periodic spinor coordinates 
and belong to the same super multiplet as the massless vector 
states,  
are given, using the SO(8) Dirac matrices $\gamma_{A\dot{A}}^i$, as $u_{\dot{A}+}\gamma^i_{A\dot{A}}S_{0A}|i\rangle$ 
where $S_{0A}$ are the zero-mode of the spinor coordinates.  
The dependent components of the spinor wave 
function $u_{A+}$ with nonzero momentum 
are given by the Dirac equation as 
$u_{A+}=-\gamma_{A\dot{A}}^ik^i u_{\dot{A}}/k^{+}$. 
From the 10 dimensional viewpoint, the wave functions 
$u_{A-}(k)$  as independent 
components of the spinor wave function have the opposite 
chiralities to $u_{\dot{A}+}(k)$ of even fermionic 
state.  Thus we can now make a consistent 
interpretation  that 
the usual GS string spectrum with periodic boundary 
condition and the new GS string spectrum with 
antiperiodic boundary condition corresponds to 
ordinarily and to oppositely GSO projected 
sectors, respectively,  in the NSR formalism.

The conservation of supercharges is 
violated at the open-string boundary $\sigma=\pi$ 
for the opposite boundary condition.   
\EQA
{d Q_A(\tau) \over d\tau}&=&-2(2p^{+})^{1/2}S_A(\tau-\pi)  ,\\
{d Q_{\dot{A}}(\tau) \over d\tau} 
&=& -2(p^+)^{-1/2}\gamma^i_{A\dot{A}}p^i(\tau-\pi) S_A(\tau-\pi)  .
\EQN
The right hand side of these equations, however,  coincide with  the zero-momentum limit of the fermion emission 
vertices of GS open string theory, just the same phenomena 
as in the NSR formalism. 
Note that the world-sheet conformal dimensions  
of these vertex operator are not equal to one. 
This is allowed in light-cone gauge, since 
we have already chosen a preferred conformal 
frame.  In the  covariant notation, 
the vertex operators at zero-momentum 
appearing in the right-hand side would be 
$p_{\mu}(z)\gamma^{\mu}_{\alpha\beta}\theta^{\beta}(z)$ 
(apart from the overall normalization factor $S_A/\sqrt{p^+}
\rightarrow \theta^{\beta} $) 
which has the correct transformation property as 
the source of a space-time massless spinor field,  
since the spinor  coordinate is a world-sheet scalar 
before $\kappa$-gauge fixing. 
Since the fermion emitted  by these  
vertices are the zero-momentum limit of 
the massless fermion state with the 
ordinary GSO projection (even fermionic G-parity), 
the form of the vertices coincides with that 
of the ordinary GS strings.  The G-parity of 
the states does not change with the insertion 
of the corresponding vertices. 
  
The next question is then how to describe the 
emission of states appeared in the opposite GS string, 
namely, the open string states with odd (both 
bosons and fermions) G-parity.  
It is clear that the corresponding vertices 
have to interchange the states between 
periodic and antiperiodic boundary conditions for 
spinor coordinates $S_A(\tau, \sigma)$.   
In the case of the NSR formalism, the change of 
boundary condition was required to describe the 
emission of space-time fermions.  For the emission of the 
oppositely GSO projected states, the boundary condition 
was preserved.  The situation in the 
GS formalism is thus just reversed: 
the change of the boundary condition is necessary 
for describing the emission of odd G-parity states, 
not for the reason of the emission of space-time fermions.   
To achieve this, 
we have to introduce an operator which flips 
the boundary condition of the spinor coordinate, an analogue 
of the spin operator in the NSR formalism 
which changes the boundary condition 
of the vector Neveu-Schwarz coordinate.  
In contrast to the spin operator, 
it is basically a space-time boson, 
which we call the the `vector operator', 
since it necessarily transforms as a space-time vector. 

As is well known\footnote{
See {\it e. g.} the chapter 5 of \cite{gsw}.  }, the `triality' symmetry of the 
SO(8) group makes possible to establish 
a direct relation between the spinor and the vector 
representations through the bosonization of the spinor coordinate.  
This is achieved by introducing 4 scalar coordinate 
$\phi_1, \ldots, \phi_4$ with the correlator 
$\langle \phi_i(z)\phi_j(w)\rangle =
-\delta_{ij}\ln (z-w)$. The eight components of the spinor 
coordinates are given as 
\EQ
\{S_A\}=\{\e^{i{1\over 2}(\pm\phi_1\pm\phi_2\pm\phi_3\pm\phi_4)}\} ,
\EN
where the number of minus sign is even.\footnote
{Here and in what follows we suppress the 
cocyle factors. } 
Then the transverse vector operator $\psi_i \, (i=1, \dots, 8)$ can be defined by 
\EQ
\psi_{2i-1}=\sqrt{{1\over2}}(\e^{i\phi_i}+ \e^{-i\phi_i}) ,
\quad
\psi_{2i}=-i\sqrt{{1\over 2}}(\e^{i\phi_i}-\e^{-i\phi_i}) .
\label{vecop}
\EN
We have used the same symbol $\psi_i$  for the 
vector operator as the Neveu-Schwarz coordinates in the NSR 
formalism, since they can 
indeed be identified. Conversely, the 
spin operator in the light-cone gauge 
NSR formalism can be identified with the 
spinor coordinate of the GS formalism. 
Note that both of the spinor coordinate 
and the vector operator have conformal dimension $1/2$. 
It is convenient to introduce also the spinor coordinates 
$S_{\dot{A}}$ in 
the ${\bf 8}_c$ representation by 
\EQ
\{S_{\dot{A}}\}=\{\e^{i{1\over 2}(\pm\phi_1\pm\phi_2\pm\phi_3\pm\phi_4)}\} ,
\EN
where the number of minus sign is now odd. 
Then by choosing the appropriate linear basis 
for the components of the spinor coordinates such that 
$S_A$ are self-conjugate, analogously as that for the 
vector operator (\ref{vecop}),  
the singular parts  of the OPEs  are given as 
\EQA
\psi_i(z) S_A(w) &\sim& (z-w)^{-1/2}\gamma^i_{A\dot{A}}
S_{\dot{A}}(w)  ,  \nonumber \\
S_{A}(z)S_{\dot{B}}(w) &\sim& 
{1\over 2}(z-w)^{-1/2}\gamma^i_{A\dot{B}}  \psi_i(w)   ,\nonumber \\
S_A(z)S_B(w) &\sim& (z-w)^{-1}\delta_{AB}  , \\
\, \, \, S_{\dot{A}}(z)S_{\dot{B}}(w) &\sim& (z-w)^{-1}\delta_{\dot{A}\dot{B}}  , \nonumber \\
\psi_i(z)\psi_j(w) &\sim& (z-w)^{-1} \delta_{ij} , \quad etc.  \nonumber
\EQN
These OPEs show that the vector operator $\psi_i$ indeed 
changes the boundary condition of the spinor coordinates 
between periodic and anti-periodic.  It should also be 
noted that with the insertion of the vector operator 
the spinor coordinate $S_A$ effectively behave as 
$S_{\dot{A}}$ which has the opposite chirality and hence 
directly couples with the independent component $u_{A-}$ 
of the massless fermion state of odd fermionic G-parity.  

We can now write down the vertex operators 
which describe the emission of the odd G-parity 
states from open GS strings, to be inserted 
at the string end points.  For example the tachyon 
vertex operator is $k_i \psi_i\, \e^{ik_ix^i}$, assuming that 
the momentum is in the transverse direction. 
The fermion emission vertex operators for odd fermionic 
G-parity are obtained 
by interchanging the role of the original spinor 
coordinate $S_A$ by $S_{\dot{A}}$ in the 
vertex operators  for the ordinary GS string theory. 
In particular, their 
zero-momentum limits are 
$
S_{\dot{A}}$ and $ \gamma^i_{A\dot{B}}p^i S_{\dot{B}}(z) $.  
Thus the corresponding supercharges and currents are 
\EQA
\tilde{Q}_{\dot{A}}(\tau)&=&(2p^+)^{1/2}\int_0^{\pi} d\sigma
(S_{\dot{A}}(\tau-\sigma)+S_{\dot{A}}(\tau+\sigma)) ,\\
\tilde{Q}_{A}(\tau) &=&(p^+)^{-1/2}\gamma^i_{A\dot{A}}
\int_0^{\pi} d\sigma(p^i(\tau-\sigma)S_{\dot{A}}(\tau-\sigma) 
+p^i(\tau+\sigma)S_{\dot{A}}(\tau+\sigma))
\EQN
\EQA
s^0_{\dot{A}}(\tau)&=&(2p^+)^{1/2} 
(S_{\dot{A}}(\tau-\sigma)+S_{\dot{A}}(\tau+\sigma)) ,\\
s^1_{\dot{A}}(\tau)&=&(2p^+)^{1/2} 
(S_{\dot{A}}(\tau-\sigma)-S_{\dot{A}}(\tau+\sigma)) ,\\
s^0_{A}(\tau) &=&(p^+)^{-1/2}\gamma^i_{A\dot{A}}
 (p^i(\tau-\sigma)S_{\dot{A}}(\tau-\sigma) 
+p^i(\tau+\sigma)S_{\dot{A}}(\tau+\sigma)) , \\
s^1_{A}(\tau) &=&(p^+)^{-1/2}\gamma^i_{A\dot{A}}
 (p^i(\tau-\sigma)S_{\dot{A}}(\tau-\sigma) 
-p^i(\tau+\sigma)S_{\dot{A}}(\tau+\sigma)) .
\EQN
The spinor coordinate $S_{\dot{A}}$ is antiperiodic 
if and only if  it  acts either on the 
fermionic states of even G-parity 
(namely, the usual GS string) or on the bosonic states 
of odd G-parity, as exhibited in the OPEs.
These properties are  identical with the 
results in the NSR formalism.  
Thus we can interpret the 32-components supercharges 
$(Q_A, Q_{\dot{A}}, \tilde{Q}_{\dot{A}} 
,\tilde{Q}_A)$ of the extended GS formalism 
as the light-cone version of 
 the $N=2$ (IIA) supercharge $Q_{\alpha}$ 
of the covariant NSR formalism. 

\section{Supersymmetry transformation law for string fields}
\setcounter{equation}{0}
The foregoing arguments provide a strong support for the 
existence of nonlinear  $N=2$ space-time 
symmetry in open superstring theory 
without the GSO projection. However, it is based on the standard world-sheet 
picture which is only first quantized. 
As an attempt toward a non-perturbative 
formulation, 
let us now turn to the discussion of space-time supersymmetry 
in the framework of string field theory.  
Although string-field theories  are 
problematic with respect to the question whether they really 
provide a well-defined theory for nonperturbative 
physics in general, we can at least formally check 
that the mechanism found in the world-sheet picture leads to the correct behaviors for general multi-body states. 
 
A natural  starting point in this direction is 
 Witten's open string field theory 
\cite{wittensft}\cite{sft1}.  
The action for the string field $A=(a, \psi)$ where 
$a$ and $\psi$ are string fields for NS and R sector, 
respectively, 
can be expressed, using the original notation of 
ref. \cite{sft1}, as 
\EQ
S={1\over 2}\int (a \ast Q_B a +   \psi \ast Y Q_B \psi + 
{2\over 3}a \ast X (a \ast a) + 
2\psi \ast a \ast \psi)  . 
\label{wittenaction}
\EN
The $Q_B$ is the BRST operator of the NSR open strings. 
The ghost number of the string fields is 
gh($a, \psi) =(-1/2, 0)$. 
The star product $\ast$ is a half-sewing operation of strings,  
which has ghost number $3/2-1=1/2$. The integral $\int$ 
is a self-sewing operation with ghost number 
$-3/2+1=-1/2$.  The BRST operator 
acts on the star products of the string fields 
as a graded derivation operator. 
The operators $X$ and $Y$ are 
the picture changing and inverse picture 
changing operators, respectively, satisfying 
$XY=YX=1$, 
 inserted at the string midpoint $\sigma=\pi/2$. 
Their  ghost numbers are $1$ and $-1$, respectively. 
If we bosonize the ghost fields appropriately, 
we can easily write down concrete expressions 
as the operators on string fields. 
The peculiar ghost numbers assigned 
to the definitions of the product and integral,  explained by the ghost current anomaly, are 
just necessary to cancel the midpoint anomalies 
of the vertices against BRST and reparametrization 
symmetries, as demonstrated explicitly 
using the oscillator representations in several 
 works \cite{sft2}.  
The gauge transformation is 
defined to be 
\EQA
\delta a &=& Q_B\epsilon + X(a\ast \lambda -
\lambda \ast a) +\psi\ast \chi -\chi \ast \psi 
\label{gaugetrans1}\\ 
\delta \psi &=& Q_B \chi + 
X(a\ast \chi - \chi \ast a) + X(\psi \ast \lambda 
-\lambda \ast \psi) 
\label{gaugetrans2} 
\EQN
Note that the gauge parameter fields  $\Lambda =(\lambda, \chi)$ have 
ghost number $(-3/2,-1)$.  
Following Witten, it is convenient to 
introduce the following  symbolic 
notation, 
\EQ
S={1\over 2}\oint (A\star Q_BA + {2\over 3} A\star A\star A) ,
\label{symaction}
\EN
\EQ
\delta A=Q_B A +A\star \Lambda -\Lambda \star A
\label{symgauge} .
\EN

For establishing gauge symmetry, the cyclic symmetry 
property of the integral of the $\ast$ products is 
essential 
\EQ
\oint A\star B =(-1)^{|A||B|} \oint B\star A ,
\label{cyclicsym}
\EN
where $(-1)^{|A|}=\pm 1$ is the grading of the string fields 
which is defined by combining the 
space-time boson-fermion  grading 
and the world-sheet grading. Thus it is necessary that 
all the string fields have definite gradings. 
In the case of bosonic string, we define the 
string fields with ghost number $-1/2$ are of odd grading 
$|A|=1$. 
The odd grading is chosen to ensure the existence 
of the kinetic term :  Since the BRST operator is 
odd, the cyclic symmetry of the integral leads to $\oint A\star Q_BA=\oint (Q_BA)\star A$ 
which must be consistent with 
$0=\oint Q_B(A\star A)=\oint ((Q_BA)\star A+(-1)^{|A|}A\star  
Q_BA)$ corresponding to the derivation property 
of the BRST operator. This requires that the 
string field $A$ has odd grading.  
Then, all the string fields of ghost number 
$2n - 1/2$ have odd grading and those 
with ghost number $2n+1/2$ have 
even grading. For example, the gauge parameter 
with ghost number $-3/2$ is even, and hence 
gauge invariance follows. We note that the integral 
is nonvanishing only if the total grading of the 
integrand is odd. Thus in the cyclic symmetry 
property  (\ref{cyclicsym}), $(-1)^{|A||B|}=1$. 

For the NSR strings, we again require that both the 
NS and R fields have odd grading for the existence of 
the kinetic term. Thus  
the gauge parameters $\epsilon, \chi$ are even.  
In other words, since the 
coefficient fields for them are respectively 
space-time bosons and 
fermions, this amounts to requiring that the world-sheet states 
have opposite but definite world-sheet gradings, odd and even, respectively. 
Then, for the NS fields with fixed total ghost number $-1/2$ 
we have to impose that $(-1)^{N_{\psi}+N_{\gamma-\beta} }$ have always the same sign for the 
excitations of the world sheet spinor field $\psi_{\mu} (\tau,\sigma) $ and the superghost fields 
$\beta(\tau, \sigma), \gamma(\tau, \sigma)$.  
For the R fields with ghost number $0$ and 
whose world-sheet grading is defined by 
$(-1)^{N'_{\psi}}\gamma_{11}$ where $N_{\psi}'$ 
is the number operator of the world-sheet spinor 
excluding the zero-mode,  
we have to require that $(-1)^{N'_{\psi}-N_{\gamma-\beta}}\gamma_{11}$ 
has the same sign for all the components. 
These conditions are satisfied if we make the ordinary GSO projection 
$(-1)^{N_{\psi}+N_{\gamma-\beta} }=1 $ and 
$(-1)^{N'_{\psi}-N_{\gamma-\beta}}\gamma_{11}=1$, 
for the NS and R fields respectively.  

This is the standard formulation of the open superstring 
field theory. 
 Actually, there is a potential 
problem in these formal arguments. Namely, 
the mid-point insertion of picture changing operator 
causes a problem in proving the 
gauge invariance to order $g^2$, since the product 
of $X$ at the same world-sheet position is singular. 
For the moment, we neglect this problem and try to 
extend the construction to include the 
string fields with opposite GSO projection. 

Let us denote the string field with the ordinary 
GSO projection by $(a_+, \psi_+)$ and those with 
the opposite projection by $(a_-, \psi_-)$, respectively.  
The above discussion shows an immediate difficulty 
in including the fields with odd G-parity. 
Namely, they obviously 
have opposite relative gradings 
compared with the ordinary GSO projection. 
Thus the requirement that the string fields must be 
of odd grading is no more satisfied, 
{\it if} the vacuum for both case is common and hence is of the same (even) grading. 
In particular, we cannot express the 
kinetic term for the odd G-parity fields 
in Witten form. Remember however that the 
kinetic terms of the standard scalar product form 
$\langle a| Q_B|a\rangle$ exist for both gradings, assuming 
appropriate reality condition. 
To express the action in Witten form, we are forced to 
assume that the Fock vacuum on which 
we construct the odd G-parity string fields 
in terms of oscillators 
have the opposite grading compared with 
the ordinary projection with even G parity.  
Since the string fields with different 
G-parity can be regarded as being independent of 
each other, this can  be done consistently 
if we take into account the 
conservation of G-parity, by introducing  a 
G-parity coordinates $\varrho$ which is even and 
satisfies $\varrho^2 =1$.  
The integration rule is $\int_G 1=1, \int_G\varrho=0 $.  
 We assign $\varrho$ to odd G-parity 
string fields,  such that only even number of 
odd G-parity fields can appear in the action.  
Grading of the string fields is always odd, irrespectively 
of G-parity in this convention.  
Also we can consistently set the Fock vacuum obtained 
from the product of two odd G-parity vacua  is even. Of course, the product of one odd and one even 
vacua is odd. 

   We denote the extended integration operation 
using the same original notation $\int \times \int_G
\rightarrow \int$.  The extended integration 
is again nonvanishing only if the integrand is odd. 
Using this convention, the standard kinetic 
term  is 
represented  in the Witten form 
$\int A\star Q_B A = \langle a_+ |Q_B| a_+ \rangle 
+ \langle \psi_+ | YQ_B| \psi_+ \rangle  +
\langle a_- |Q_B| a_-0 \rangle 
+ \langle \psi_-| YQ_B| \psi_- \rangle$. 
Thus we can 
unify the even and odd G-parity 
fields and denote them as 
$A=(a, \psi) = ( a_++\varrho a_-,  \psi_+ +\varrho \psi_-)$. 
 which is of grading odd as a whole. Hence,  
the same form (\ref{wittenaction}) or 
(\ref{symaction}) for the action is valid, 
with gauge transformation being extended 
to both G-parities by defining the extended gauge parameter 
field $\Lambda = ( \lambda_+ + \varrho\lambda_-, 
 \chi_+ + \varrho\chi_-)$. 

We can now proceed to investigate the 
space-time supersymmetry.  
The natural candidate for the supersymmetry 
generator is  the zero-momentum 
fermion vertex operator (\ref{scharge}).  Witten's ansatz 
which has correct ghost number and grading is 
\EQ
\delta a =\epsilon^{\alpha}Q_{\alpha} \psi ,\quad \delta \psi =
X({1\over 2}\pi) \epsilon^{\alpha} Q_{\alpha}a . 
\label{sfieldsuper}
\EN 
To show the invariance of the action under 
this transformation, we need that  $Q_{\alpha}$ satisfies 
 \EQ
\oint \epsilon^{\alpha}Q_{\alpha} A=0 , 
\label{schargecons1}
\EN
\EQ
\epsilon^{\alpha} Q_{\alpha} (A\star A) = 
(\epsilon^{\alpha}Q_{\alpha}A) \star A 
+A\star \epsilon^{\alpha}Q_{\alpha}A . 
\label{schargecons2}
\EN
These are sufficient for proving the  invariance of the cubic 
interaction term. They 
are indeed satisfied, as explicitly checked in the 
oscillator representation in ref. \cite{antal} for 
the case of ordinary GSO projected formalism. 
In our case with the oppositely projected sectors, 
we have to take into account the G-parity coordinate $\varrho$ 
to ensure the correct grading and G-parity properties in establishing 
(\ref{schargecons1}) and (\ref{schargecons2}).  
The super parameter $\epsilon^{\alpha}$, which 
of course is of odd grading,  should also 
be decomposed as $\epsilon^{\alpha} =  
\epsilon_+^{\alpha} + \varrho \epsilon_-^{\alpha} $ 
where $\epsilon_{\pm}^{\alpha}$ are 
the 10 dimensional Weyl spinors with positive and negative fermionic G-parity, respectively.  Since, apart from the 
grading, only essential requirement for 
the validity of these relations is that the 
supercharge is an integral along the strings 
of the current operator which is not 
too singular with respect to the midpoint insertions. 
Since the latter property of the world-sheet supercurrent, being 
ensured by the conformal dimensions of the operators,   
is not affected by  the opposite projection, we can conclude that 
they are satisfied with the existence of both  projections as equally well 
as in the case of the ordinary projection.  
Thus the cubic terms alone 
have linear space-time supersymmetry. 

To show the supersymmetry of the quadratic 
kinetic term, it is necessary to use the fact that the supercharge 
is BRST invariant in addition to the above 
properties.  It is at this point where 
a crucial difference of  our case from the theory 
with the ordinary projection arises.  
The commutation relation of the fermion emission 
vertex and the BRST operator is a total derivative, 
\EQ
\{Q_B, V^{F}_{\alpha}(z) \} =\partial_z(c(z)V^F_{\alpha}(z)) . 
\EN
Thus supercharge is BRST invariant if and only if 
the fermion emission operator is single-valued. 
If it is double-valued, we have,  using the same convention 
for open string notations ($\tau=0$) as in subsection 4.1, 
\EQ
[Q_B, \epsilon^{\alpha}Q_{\alpha}] =-
c \epsilon^{\alpha}V^F_{\alpha}(\pi) +c \epsilon^{\alpha}V^F_{\alpha}(-\pi)=-2c\epsilon^{\alpha}
V^F_{\alpha}(\pi) .
\EN
This is just the manifestation of the same phenomena which 
we  have been discussing  
in previous sections on the basis of the first quantized world-sheet picture. 
In terms of string field theory, 
the insertion of fermion emission vertices must equivalently 
be described by an appropriate shift of the string field. 

The variation of the quadratic term 
under the linear super transformation is 
\EQ
\delta \oint A\star Q_B A = 2\int \psi\ast [ Q_B, 
\epsilon^{\alpha}Q_{\alpha}]a  .
\EN
The right-hand side is nonvanishing if and only if 
the NS string field $a$ is in the odd G-parity sector 
and hence the fermionic G-parity of 
the R field $\psi$ is opposite to that of the 
parameter of supersymmetry transformation. 
The result is of course identical with world-sheet 
approach. 
Here we have used the fact that the picture 
changing operators satisfy $XY(\pi/2)=1$  and 
commutative with both BRST operator and 
the supercharge.  
The commutativity of the picture changing 
operator with the supercharge is valid 
even if the fermion emission vertex is 
double-valued since it commutes with the 
ghost variable $\xi$ : $[Q_{\alpha}, X] = 
[Q_{\alpha}, \{Q_B, \xi\}]=\{2cV^F(\pi), \xi\}=0$. 
This violation of BRST invariance 
 is canceled if we shift the fermionic string field as 
\EQ
\delta_{shift} \psi = \epsilon^{\alpha}Q_{\alpha}Q_L{\cal I} ,
\EN
where $Q_L$ (and $Q_R$) is the `half' BRST operator 
which is defined by integrating over the left (right) half 
of the open string $Q_B=Q_R + Q_L$.  Also we have 
introduced the `identity' string field ${\cal I}$ as introduced in \cite{hlrs}, 
which is  characterized by 
\EQ
{\cal I}\star A =A\star {\cal I}=A , \quad  \forall A
\EN
and has ghost number $-1/2$, due to the midpoint insertion. 
Note that the string field $Q_L {\cal I}$ is odd.  
Some other useful properties of these operators 
are 
\EQA
Q_B {\cal I} &=& 0 , \\
Q_{\alpha}{\cal I} &=&0 , \\
(Q_R A)\star B &=& -(-1)^{|A|}A\star Q_LB , \\
\{Q_B, Q_L\}&=&\{Q_B, Q_R\}=0 . 
\EQN
Using these relations we derive 
\EQ
\delta_{shift} \int \psi\ast a \ast \psi = - \int \psi \ast [Q_B, \epsilon^{\alpha}Q_{\alpha}] a  ,
\EN
\EQ
\delta_{shift}\int \psi \ast Y Q_B \psi = -\int 
Q_B Y\epsilon^{\alpha}Q_{\alpha}Q_B\psi =0 .
\EN
This shows  that the total action (\ref{wittenaction}) 
 is invariant under the 
nonlinear supersymmetry transformation obtained by 
combining (\ref{sfieldsuper}) and $\delta_{shift}\psi$, 
\EQA
\delta a &=& \epsilon^{\alpha}Q_{\alpha} \psi , \\
\delta \psi &=&  
\epsilon^{\alpha}Q_{\alpha}Q_L{\cal I}+
X({\pi\over 2}) \epsilon^{\alpha} Q_{\alpha}a .
\label{sfieldpsi}
\EQN

We  recognize that 
the first term in (\ref{sfieldpsi}) can actually be absorbed into 
the second term by making the following field redefinition, 
\EQ
a + Y({\pi\over 2})Q_I {\cal I} \rightarrow a  
\label{fieldredef}
\EN
which simultaneously eliminates the quadratic terms 
from the action. 
This property is a direct consequence of the 
fact that the string field action can be 
derived from the action with only the 
purely cubic interaction terms \cite{hlrs} by 
making the field redefinition 
$a \rightarrow a + Y({\pi\over 2})Q_I {\cal I}$. 
Therefore the above nonlinear (inhomogeous) transformation 
law is simply a corollary of the supersymmetry of the cubic 
interaction terms under the linear supersymmetry 
transformation.  

In this way, 
all the results of the present section nicely fit to what we have 
derived in the world-sheet approach.    
The existence of 
purely cubic action is a natural consequence of the general 
fact that in the 
world-sheet approach, especially in its covariant 
version, the free motion and the interactions of 
strings are one and the same thing from the viewpoint of the 
local world-sheet dynamics. In fact, that was the 
original motivation for the cubic action conjecture 
\cite{cubic}\cite{hlrs}\cite{kyoto}.

\section{Discussions}
\setcounter{equation}{0}
\noindent
{\it Problems of contact terms}

\noindent
As we have already mentioned during the discussions 
of string-field theory approach, there is 
the  subtle question of possible contact terms \cite{wendt} 
beyond 3-point interactions. This is basically 
due to  singularities caused by the 
midpoint insertion of picture changing 
operators.  For a recent discussion on this 
problem, see ref. \cite{berk} and references therein. 
If the higher contact terms are really inescapable, 
the formal transformation law given in section 5 
might also be subject to corrections which 
are at least quadratic in string fields. 
The investigation along this direction is 
beyond the scope of the present paper. 
Our arguments using the ordinary world-sheet 
approach  already provided reasonable evidence that there 
exists the $N=2$ space-time supersymmetry. 
The problem of contact terms might be an 
indication that we should seek for 
other more tractable approach 
to formulate the nonperturbative 
physics of string theory than the traditional 
string-field theory approach. 

\noindent
{\it Chan-Paton factor}

\noindent
It is straightforward to include the Chan-Paton factor. 
Let us only mention the case of string field theory. 
For unstable $N$ D9-branes of type IIA theory, 
the GSO projection is not done at all and hence 
we  simply extend all the string fields 
to be hermitian $N\times N$ matrices which transform 
in the adjoint representation of U($N$) group,   
\[
a \rightarrow   a_{+,ij}+\varrho a_{-,ij} ,\quad 
\psi \rightarrow  \psi_{+, ij} + \varrho \psi_{-, ij}  . 
\] 
The star multiplication of the string fields is supplemented 
by the matrix multiplication and the 
integration by the trace operation. Note that the supersymmetry 
parameter must be singlet under the gauge group. 
Corresponding to the existence of both G-parity fields, 
we have IIA type supersymmetry. 

In type IIB theory, we consider the systems with equal 
number of D9 and anti-D9 branes 
to cancel the tadpole anomaly coming from the 
RR 10 form potential. For  strings  
which connect D9-branes only within the 
D9-branes or only within the anti D9-branes, 
the ordinary GSO projection is assumed, 
while for strings which connect D9 and anti-D9 branes 
the opposite GSO projection is assumed. 
Therefore, the inclusion of Chan-Paton factor is 
represented by the matrix string fields of the following form,  
which transform as the adjoint representation of 
U($N$)$\times$U($N$) group, 
\[
a \rightarrow \pmatrix{
 a_{+, ij} & \varrho \, a_{-, ik} \cr
\varrho \, a^{\dagger}_{-, \ell j} &  \tilde{a}_{+, \ell k} \cr}
, \quad 
\psi \rightarrow \pmatrix{ 
\psi_{+, ij} & \varrho \, \psi_{-, ik} \cr
\varrho \, \psi^{\dagger}_{-, \ell j} &  \tilde{\psi}_{+, \ell k} \cr} .
\]
Actually, our discussions in previous sections must be subject to a small change in type IIB case.  As shown in the 
above assignment of the opposite G-parity fields, the 
transformation laws of the string fields with respect to 
the group U($N$)$\times$ U($N$) are  different 
for the positive and negative G-parity fields. 
To be consisitent with the gauge symmetry, the 
supersymmetry transformation must be 
singlet under the gauge group. This forbids the 
supersymmetry parameter with negative G-parity. 
However, there are now two independent 
supersymmetry parameters $\epsilon_+$ and 
$\tilde{\epsilon}_+$ , both of which belong to 
the same positive chirality corresponding to 
the zero-momentum limit of the two positive G-parity 
open string fields $\psi_{+,ij}$ and $\tilde{\psi}_{+,\ell k}$. 
This leads to a chiral $N=2$ supersymmetry as it should be.

\noindent
{\it Condensation of tachyon}

\noindent
An  important future problem is to investigate the 
tachyon condensation in light of 
our arguments.  There is already a 
discussion by Sen \cite{sen2}(see also a more 
recent work by Sen and Zwiebach \cite{senzwie} based on 
some old works \cite{samuel}) of tachyon condensation 
from the viewpoint of string 
field theory. We do not have much to say in the 
present paper other than what we have already 
remarked. However, let us look at the concrete form of the 
field equation which we have to solve. 
It is sufficient to  study the problem 
after making the field redefinition (\ref{fieldredef}). 
The field equations are then, neglecting the 
possible higher contact terms and integrating over the 
G-parity coorinates,  
\[
a_+\ast a_+ + a_-\ast a_- =0, 
\]
\[
 a_-\ast a_+ + a_+ \ast a_-=0 ,
\]
\[
a_+\ast \psi_+ + \psi_+ \ast a_+ +
a_-\ast \psi_- + \psi_- \ast a_- 
=0 ,
\]
\[
a_-\ast \psi_+ + \psi_- \ast a_+ +
a_+\ast \psi_- + \psi_+\ast a_- =0 .
\]
The condensation of 
tachyon corresponds to nonzero $a_-$. 
Also, the supersymmetric (and/or  BPS) solutions  
with vanishing fermion backgrounds must satisfy 
\[
\epsilon^{\alpha}Q_{\alpha} a = 0
\]
for certain constant spinor $\epsilon^{\alpha}$. 
It is important to note that the solutions described by these 
equations do not have the ordinary 
kinetic term in the action.  Thus even if 
some of the gauge symmetry remains unbroken after 
the tachyon condensation, there is in general no 
propagation of massless fields corresponding 
to unbroken gauge symmetry.  Instead, the 
propagating modes must be 
determined anew on the new background defined 
by the solutions themselves. This conforms to 
some related observations made in \cite{sen1}\cite{yi}  
and seems to indicate the resolution of 
a puzzle \cite{witten} associated with the K-theoretical 
interpretation of D-branes.  
Remember also that this phenomenon has 
already been anticipated in a 
motivating discussion in  section 2.  
The  condensation  of tachyon makes the 
supersymmetry transformation linear and 
simultaneously eliminates the kinetic term. 
It is interesting to investigate how to 
represent this situation in a low-energy 
field-theory approximation.  
Work along this direction is in progress and will be
reported in a forthcoming paper \cite{hy}. 
 
Finally, one of the  fundamental problems related to 
the subject of the present work is to establish 
connections with the ordinary closed-string approach of 
type II string theory. That would 
not only lead to 
a new unexpected connection among different field theories 
in the low-energy limit, but also provide an important clue 
towards the dynamical formulation of M-theory.   

\vspace{0.2cm}
\noindent 
Acknowledgements

The author would like to thank his colleagues, especially 
M. Kato and Y. Kazama, for stimulating conversations  
related to the present work. 
The present work  is supported in part 
by Grant-in-Aid for Scientific  Research (No. 09640337) 
and Grant-in-Aid for International Scientific Research 
(Joint Research, No. 10044061) from the Ministry of  Education, Science and Culture.

\end{document}